\documentclass{cjaa}    
\usepackage{graphicx}                   

\begin{document}

\title{The Parkes Pulsar Timing Array}

\setcounter{page}{1}          

\author{R. N. Manchester
\inst{}\mailto{}
      }

\institute{Australia Telescope National Facility, CSIRO, PO Box 76, Epping NSW 1710, Australia\\
\email{dick.manchester@csiro.au}
          }

\date{Received~~2005 month day; accepted~~2005~~month day}

\abstract{Given sufficient sensitivity, pulsar timing observations can
make a direct detection of gravitational waves passing over the
Earth. Pulsar timing is most sensitive to gravitational waves
with frequencies in the nanoHertz region, with the most likely
astronomical sources being binary super-massive black holes in galaxy
cores.  The Parkes Pulsar Timing Array project uses the Parkes 64-m
radio telescope to make precision timing observations of a sample of
about 20 millisecond pulsars with a principal goal of making a direct
detection of gravitational waves. Observations commenced about one
year ago and so far sub-microsecond timing residuals have been
achieved for more than half of these pulsars. New receiver and
software systems are being developed with the aim of reducing these
residuals to the level believed necessary for a positive detection of
gravitational waves.\keywords{pulsars: general --- gravitational waves
--- time} }
 
\authorrunning{R. N. Manchester}            
\titlerunning{The Parkes Pulsar Timing Array}  

\maketitle

%
%
\section{Introduction}           
Gravitational waves (GW) are a prediction of Einstein's general theory of
relativity (and other relativistic theories of gravity) and are
generated by massive objects in accelerated motion relative to an
inertial frame. They are incredibly weak and, despite much effort over
many decades, they have never been directly detected. Pulsars have
provided strong evidence for their existence through observations of
the decay of the orbit of binary pulsars, in particular for the
Hulse-Taylor binary PSR B1913+16 (\cite{tw82,wt05}). Current projects
aimed at detecting gravitational waves include LIGO (\cite{aaa+04}),
VIRGO (\cite{aaa+04g}) and LISA (\cite{dan00}). All of these are laser
interferometer systems. LIGO and VIRGO are ground-based and are most
sensitive to GW at frequencies around 100 Hz, whereas the proposed LISA
system consists of three spacecraft in orbit about the Sun and is most
sensitive to GW with frequencies of about 1 mHz. 

Gravitational waves in our Galaxy modulate the apparent pulse period
of pulsars. Pulsar timing experiments measure pulse times of arrival
(TOAs) and so they are sensitive to accumulated pulse phase
offsets. Consquently, they are sensitive to long-period GW with
highest sensitivity to waves which have a period comparable to the
length of the data span. This is typically several years, so maximum
sensivity is achieved for GW with frequencies of about 1 nanoHertz.  Pulsar
timing experiments are therefore quite complementary to the
laser-interferometer systems.

Observations of a single pulsar can be used to set limits on the
strength of the GW background in the Galaxy at these nanoHertz
frequencies. Any GW signal must be smaller than the timing residuals
remaining after fitting for the pulsar model (assuming that the model
terms do not absorb the GW signal). Because of their narrow pulse
profiles (expressed in time units) and exceptional period stability,
millisecond pulsars (MSPs) give the best limits. From an 8-year
sequence of Arecibo observations of PSR B1855+09, an upper limit on
the energy density of the GW background relative to the closure
density of the Universe, $\Omega_g \sim 10^{-7}$, was obtained by
Kaspi, Taylor \& Ryba (1994).\nocite{ktr94} A longer but non-uniform
data set on the same pulsar was analysed by Lommen et
al. (2002),\nocite{lbsn03} reducing this limit by about an order of
magnitude (see also Damour \& Vilenkin 2005)\nocite{dv05}.

In principle, observations of many pulsars spread across the celestial
sphere, a ``pulsar timing array'', can make a direct detection of the
GW background at the Earth (\cite{hd83,fb90}). GW passing over the pulsars
result in period perturbations which are uncorrelated between the
different pulsars. However, GW passing over the Earth produce
correlated residuals which contain the quadrupolar signature of GW as
a function of position on the sky. In contrast, clock errors produce a
correlated signal which is independent of sky position and hence can
be separated from the GW signature. Similarly, an error in the
Solar-System ephemeris (effectively an error in the velocity of the
Earth with respect to the Solar-System barycentre) produces a dipole
signature which can be separately identified.

The Parkes Pulsar Timing Array (PPTA) project is a collaborative
effort between Swinburne University of Technology, the University of
Texas at Brownsville and the Australia Telescope National Facility to
make precision timing observations of a sample of about 20 MSPs using
the Parkes 64-m radio telescope with the ultimate goal of making a
direct detection of gravitational waves. In \S\ref{s:goals} we give a
detailed description of the goals and methods for the project,
\S\ref{s:time} discusses the prospects of defining a timescale based
on pulsars, \S\ref{s:rfi} describes our efforts at countering the
effects of radio-frequency interference (RFI) and the current state of
the project is summarised in \S\ref{s:status}.

\section{Goals and methods}\label{s:goals}
The main goals of the the PPTA project are as follows:
\begin{list}{$\bullet$}{}
\item Detection of gravitational waves from astronomical sources
\item To establish a pulsar-based timescale
\item To investigate possible errors in the Solar-System ephemeris
\end{list}
To achieve these overall goals, a large number of secondary or
intermediate goals have to be met. These include:
\begin{list}{$\bullet$}{}
\item Develop new instrumentation for precision pulsar timing at the
  Parkes radio telescope
\item Make timing observations of $\sim 20$ MSPs at 2 -- 3 week
  intervals for five or more years at three frequencies: 700 MHz, 1400 MHz and
  3100 MHz
\item Achieve daily pulse TOA precisions of $\la 100$~ns for more
  than 10 MSPs and $<1 \mu$s for the rest of the sample
\item Investigate and model the effect of GW signals on pulsar timing
  data
\item Develop software for analysis of timing data from multiple
  pulsars with systematic errors of $<2$~ns
\item Develop and implement methods for detection of GW signals 
\item Develop and implement methods for investigating instabilities in the
  terrestrial timescale and establishing a pulsar-based timescale
\item Develop and implement methods for investigating errors in the
  Solar-System ephemeris and improving the ephemeris
\item Investigate the effect of signal propagation through the
  interstellar medium and correct for it where possible
\item Develop and implement methods for mitigating the effects of RFI
  on pulsar timing data
\item Develop links with international groups to foster collaboration
  and coordination of pulsar timing efforts
\end{list}

For observations at 1400 MHz, the PPTA mainly uses the central beam of
the Parkes multibeam receiver (\cite{swb+96}) although the ``H-OH''
receiver is used occasionally. The multibeam receiver central beam is
a dual-channel system with orthogonal linear feeds. Each channel has a
bandwith of 256 MHz and a system equivalent flux density on cold sky
of about 30 Jy. At 700 MHz and 3100 MHz, the ``10cm/50cm''
receiver is used. This dual-frequency coaxial system with orthogonal
linear feeds at each frequency allows simultaneous observations in the
two bands. At 700 MHz the bandwidth is 64 MHz with a system equivalent
flux density of about 70 Jy, and at 3100 MHz the bandwidth is 1024 MHz
with a system equivalent flux density of about 45 Jy.

For observations at 700 MHz and 1400 MHz, data are recorded using the
CPSR2 baseband recording system (see \cite{jhb+05}) which 2-bit samples
two polarisations for each of two 64-MHz wide bands (at 700 MHz, only
one band can be used) and coherently dedisperses the data to give high
time resolution. The wider bandwidths available at 1400 MHz and 3100
MHz have been observed using the ``Wideband Correlator''. This
system, which is based on the Canaris correlator chip developed at the
National Radio Astronomy Observatory, can 2-bit sample bandwidths up
to 1024 MHz. It provides up to 1024 frequency channels on each of four
polarisation products and synchronously integrates these at the
topocentric pulsar period with up to 2048 bins per pulsar
period. 

\begin{figure}
\centerline{\includegraphics[width=120mm]{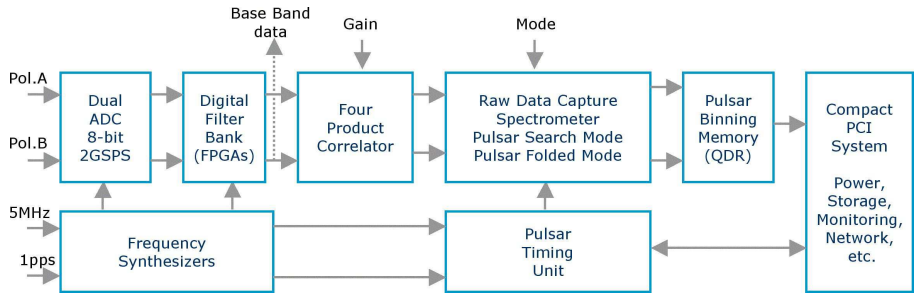}}
\caption{Block diagram of the pulsar digital filterbank system under
  development at ATNF. (G. Hampson, private communication)}\label{fg:DFB}
\end{figure}

Unfortunately, the processing speed of the wideband correlator system
allows only a limited number of channels and bins for observations of
MSPs, thereby limiting the precision of measured TOAs. To overcome
this and other limitations we are developing a pulsar digital filterbank
(PDFB) system with enhanced capabilities. This system will implement a
polyphase filter (\cite{fs04}) using Field Programmable Gate Array
(FPGA) processors and, like the wideband correlator, will have 1024
MHz maximum bandwidth. A block diagram of the system, which is being
designed and constructed by Grant Hampson and Andrew Brown at the
ATNF, is shown in Fig.~\ref{fg:DFB}. It will use dual 8-bit digitisers
operating at 2 Gsamples per second. The 8-bit digitisation provides
higher sensitivity and better fidelity in the digitised signal as well
as increased protection from strong RFI. The polyphase filter gives a
superior channel bandpass shape with about 30 db of sidelobe rejection
compared to 12 db for an FFT spectrometer, further reducing the effect
of strong RFI. The PDFB implementation will have up to 2048 frequency
channels on each of four polarisations, a pulsar binning memory with
up to 2048 bins per period and a minimum bin time of 5 $\mu$s. The
system will also have a ``search'' mode in which channel data is
dumped directly to disk at a specified sampling interval and a
baseband mode providing up to 16 baseband-sampled channels on two
polarisations, each 64 MHz wide and contiguous across the 1024 MHz
input bandwidth. These will provide the basis for a next-generation
baseband system to be developed by the ATNF in collaboration with
Swinburne University.

\begin{figure}
\centerline{\includegraphics[width=60mm,angle=270]{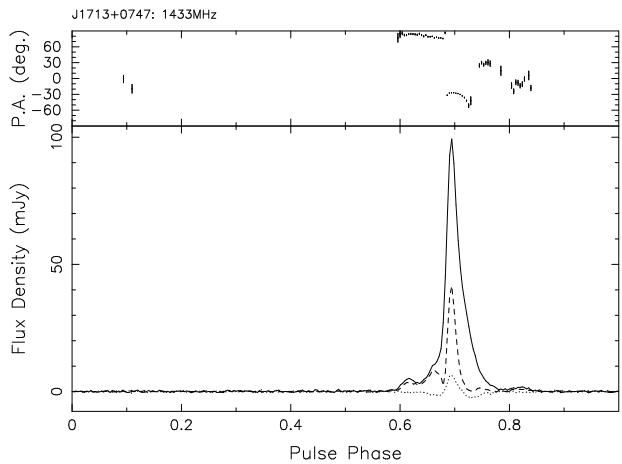}}
\caption{Polarisation of the mean pulse profile for the MSP PSR
  J1713+0747 at 1433 MHz recorded with the prototype digital
  filterbank system at Parkes. In the lower part of the figure, the
  solid line is total intensity (Stokes $I$), the dashed line is
  linearly polarised intensity ($L=(Q^2 + U^2)^{1/2}$) and the dotted
  line is circularly polarised intensity (Stokes $V$). The upper part
  shows the position angle of the linearly polarised
  component.}\label{fg:1713_poln}
\end{figure}

A prototype PDFB system which has 256 MHz maximum bandwidth was
installed at Parkes in June, 2005, and is working well. To illustrate
this, Fig.~\ref{fg:1713_poln} shows the mean pulse polarisation
profiles for PSR J1713+0747 from an observation of duration 64 min at
1433 MHz which agree well with those published by Ord et
al. (2004)\nocite{ovhb04}. This pulsar has a period of 4.57 ms and a
dispersion measure of 15.99 cm$^{-3}$ pc. The PDFB configuration used
had 512 frequency channels and 256 pulse phase bins giving a profile
resolution of 17.8 $\mu$s and a dispersion smearing of 24.2
$\mu$s. The final system will improve on these numbers, giving an
effective time resolution of about 10 $\mu$s for this pulsar.

Recording data at the telescope is only the first stage in achieving
the goals of the PPTA. Pulse profiles must be processed to remove the
effects of bad or corrupted data, calibrated to remove instrumental
effects, summed in frequency taking into account the effects of
interstellar dispersion, and summed in time to form high
signal/noise pulse profiles. These are then correlated with a standard
pulse template to give topocentric TOAs of a reference pulse phase for
each profile. The {\sc psrchive} pulsar data analysis system
(\cite{hvm04}) is used for all data processing. Once a set of TOAs for
a given pulsar have been obtained, a pulsar model is fitted to the
data to form a set of timing residuals and give improved model
parameters. 

Up to now, the program {\sc tempo}\footnote{See
http://www.atnf.csiro.au/research/pulsar/tempo.} has been widely used
for this purpose. However, {\sc tempo} contains a number of
approximations which preclude the very high precision required for the
PPTA and other similar projects. To overcome these problems and to
provide a more user-friendly and versatile interface to the program, a
new program, {\sc tempo2}, is being developed at the ATNF. This program,
described by Hobbs et al. (2005)\nocite{hem05a} in these proceedings,
will also allow simultaneous fitting of data from many pulsars, a
vital step in the process of achieving the goals of the PPTA.

\section{A pulsar timescale}\label{s:time}
The standard of terrestrial time, TT(TAI), is defined by a weighted
average of more than 200 caesium clocks at time standard laboratories
around the world. The unit of TT(TAI) is the SI second on the
rotating geoid and the rate of TT(TAI) is adjusted from time to time
to conform to this definition. TT(TAI) is defined essentially in real
time and has a fractional frequency stability of order
$10^{-15}$. With reassessment of errors in individual clocks and
improved averaging techniques, it is possible to retroactively define
an improved timescale. A number of such improved timescales have been
defined by the BIPM in Paris, e.g., TT(BIPM2003) (\cite{pet03b}). 

Soon after the discovery of the first MSPs, it was recognised that
pulsars could be used to define a timescale independent of terrestrial
clocks (e.g., \cite{tay91}). This timescale is relative in the sense
that there is no standard ``pulsar frequency'' and so can only be used
to investigate the stability of terrestrial timescales. Furthermore,
these comparisons are insensitive to linear changes in rate (since
pulsars have an intrinsic period derivative) and to annual terms since
pulsar positions are not known a priori to sufficient accuracy. From
an analysis of 7 years of Arecibo timing data, Kaspi,
Ryba \& Taylor (1994)\nocite{ktr94} showed that PSR B1855+09 has a
fractional frequency stability $\sim 2\times 10^{-15}$ averaged over
timescales of several years. 

\begin{figure*}
\begin{tabular}{ll}
\mbox{\includegraphics[width=70mm]{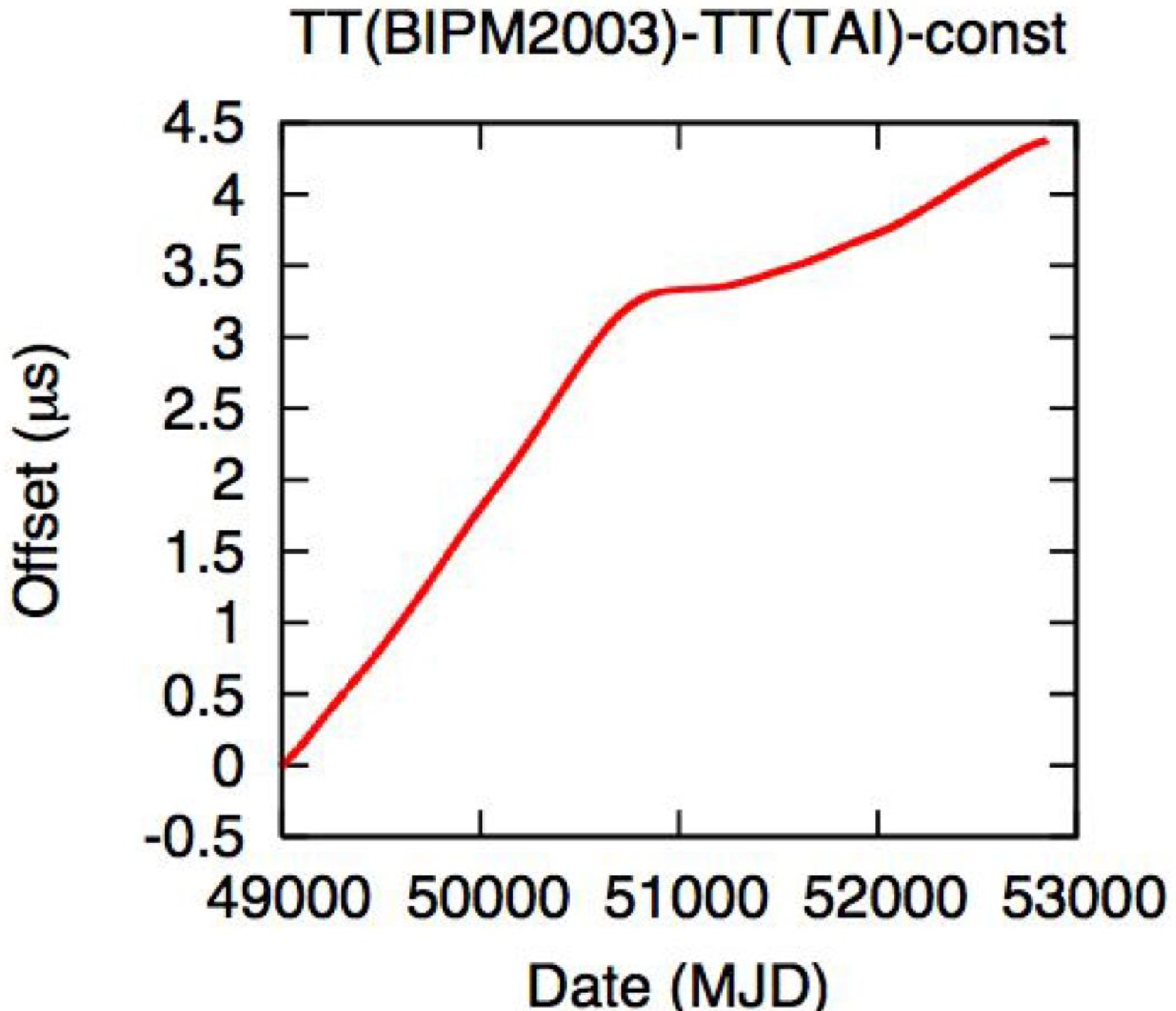}}  &
\mbox{\includegraphics[width=68mm]{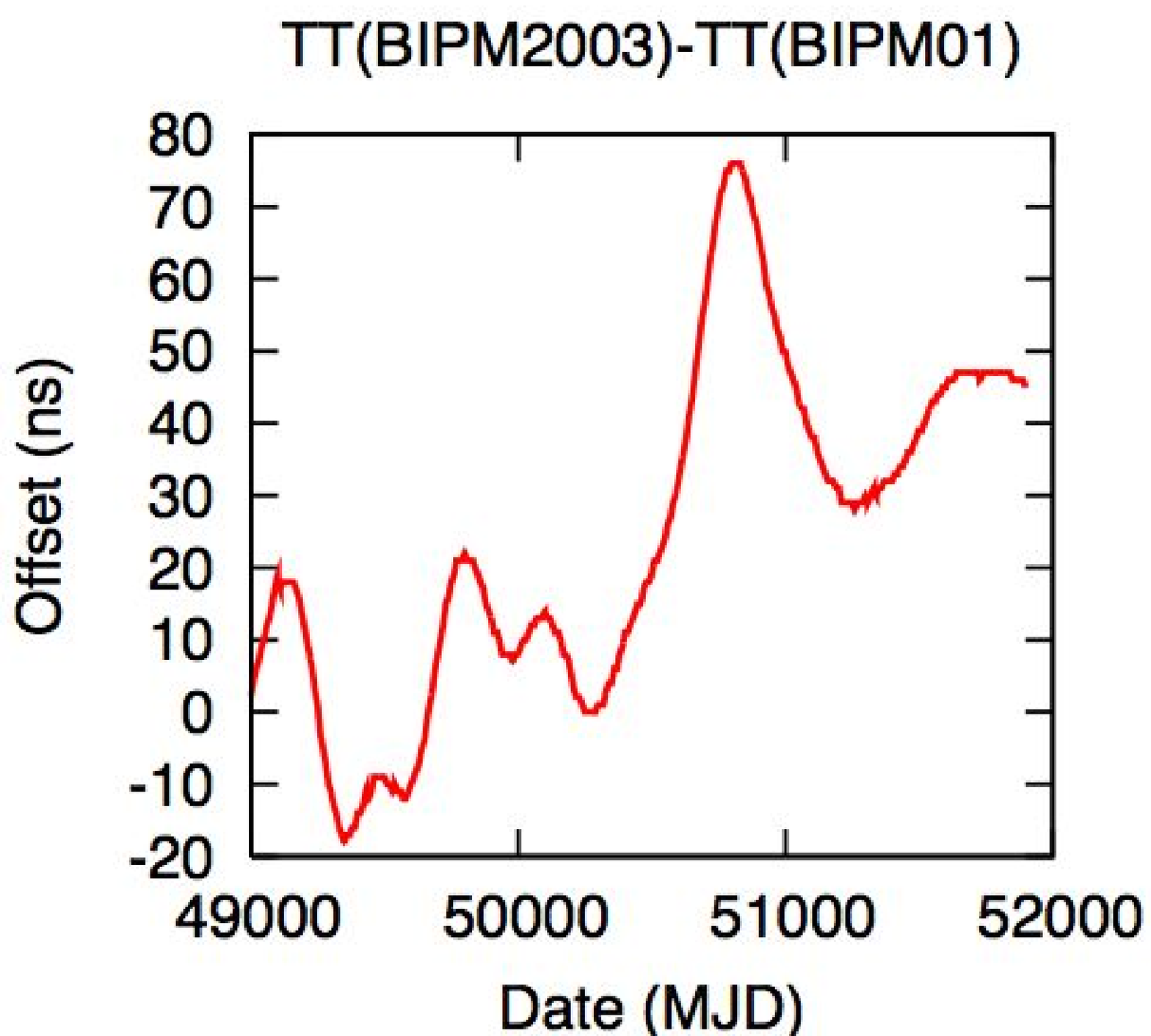}} 
\end{tabular}
\caption{(a) Difference between TT(BIPM2003) and TAI from 1993 to 2003
  after removal of a constant term. (b) Difference between
  TT(BIPM2003) and TT(BIPM2001) from 1993 to 2001. (R. Edwards with data from
  \cite{pet03b}.)  }\label{fg:time}
\end{figure*}

By averaging over many pulsars and with improved technology, timing
array observations should be able to improve significantly on this
limit and potentially detect variations in the best available
terrestrial timescale, in effect defining a ``pulsar timescale''
(\cite{fb90,pt96}). As an illustration of the magnitude of the effects
that a pulsar timescale might reveal, the left panel of
Fig.~\ref{fg:time} shows the difference between TT(BIPM2003) and
TAI. Even after the trend is removed, there are deviations of order 1
$\mu$s which will be easily detectable by the PPTA. As an indication
of the base-level uncertainty in the best available terrestrial
timescales, the right panel shows the difference between TT(BIPM2003)
and TT(BIPM01). Peak deviations are of order 50 ns. If the PPTA is
able to meet its goal of sub-100~ns timing for more than 10 MSPs, the
pulsar timescale should be defined to 10 -- 20 ns with an averaging
time of a few weeks, sufficient to detect fluctuations in the best
terrestrial timescales.

\section{RFI mitigation}\label{s:rfi}
RFI is a problem for all radio astronomy observations. Furthermore,
with increasing use of the RF spectrum and increasing sensitivity of
radio astronomy systems, it is a problem which is rapidly becoming
more serious. Because of their high time resolution and high
sensitivity, demanding wide bandwidths, pulsar observations are
especially susceptible to contamination by RFI. The problem is
especially acute for the 50~cm (700 MHz) receiver at Parkes where
digital TV transmissions have recently commenced within and adjacent
to the receiver bandpass. In this case, it is possible to obtain a
reference spectrum of the contaminating signal with high signal/noise
ratio using a 3.5-m parabolic reflector directed at the TV station
which is approximately 200 km south of the Observatory.

\begin{figure}
\centerline{\includegraphics[width=150mm]{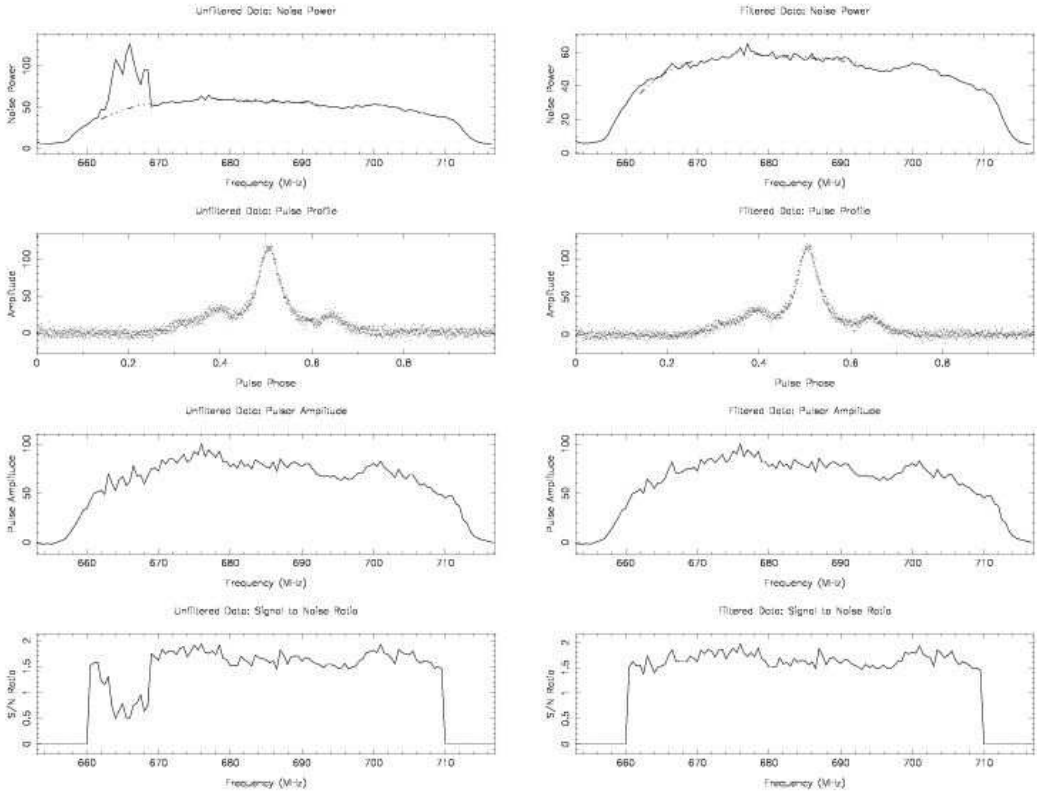}}
\caption{Results of adaptively filtering data from the 50~cm receiver
  at Parkes to remove the effects of digital TV in the bandpass. The
  left panels show properties of the signal before filtering and the
  right panels show results after filtering (\cite{khc+05}). See text
  for further details. }\label{fg:RFI}
\end{figure}

With such a reference signal, it is possible to generate an adaptive filter
to effectively remove time-varying RFI contamination from the telescope
observation without affecting the underlying astronomy signal
(\cite{bb98c,khc+05}). The results of a test of this procedure are
shown in Fig.~\ref{fg:RFI}, with the left panels showing the
unfiltered data for one receiver channel and the right panels showing
the effect of the adaptive filtering. These data are from a 16.7-s
observation of PSR J0437$-$4715 and the reference signal, both
recorded using the CPSR2 recording system. They were processed
off-line making use of {\sc psrchive}. The upper panels show the
receiver bandpass, the second panel shows the pulsar mean pulse
profile, the third panel shows the power spectrum of the pulsar signal
and the bottom panel shows the signal-to-noise ratio of the pulsar
signal. These plots clearly show that the adaptive filtering removes
the effect of the RFI without affecting the astronomy signal either
under the RFI or in the rest of the bandpass. Processing of the other
(orthogonal) receiver channel (recorded simultaneously) gave similar
results and analysis of the two channels together showed that the
polarisation properties of the pulsar signal were preserved.

Given the success of these trials, we are developing a hardware
implementation of the adaptive filter that will be able to operate in
real time for the 64-MHz bandwidth of the 50~cm receiver. This will be
based on the same FPGA processors as used in the PDFB and will act as a
preprocessor for it or other data recording systems. We expect to
commission this system during 2006. 

\section{Current status of the PPTA}\label{s:status}
\begin{figure}
\centerline{\includegraphics[width=80mm,angle=270]{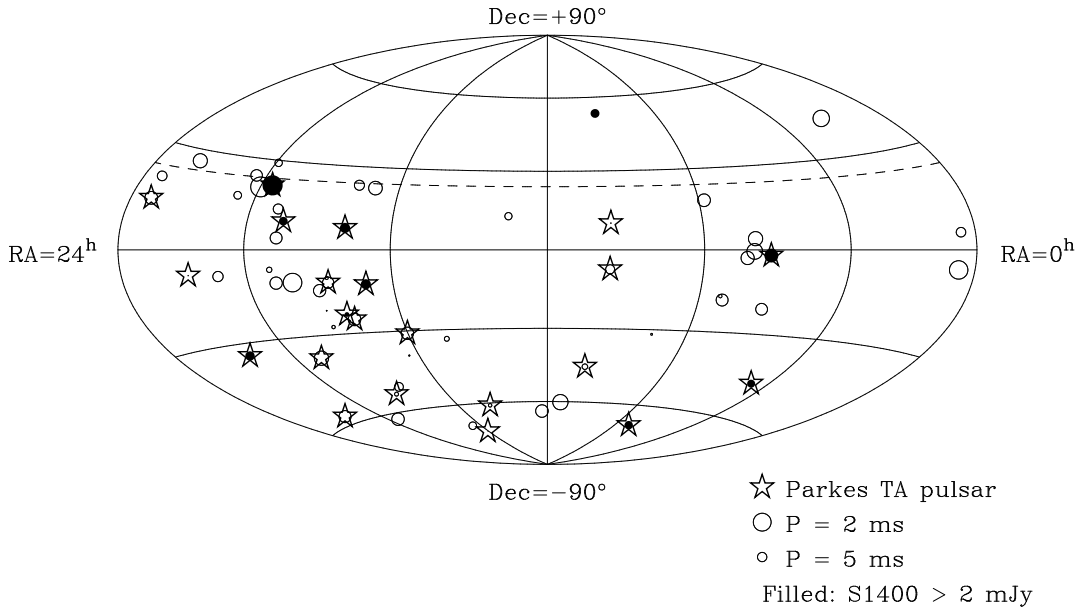}}
\caption{Galactic disk MSPs with periods less than 20 ms plotted in
  celestial coordinates. The size of the circle is inversely related
  to the pulsar period and for stronger pulsars the circle is
  filled. The dashed line is the northern declination limit of the
  Parkes telescope. Pulsars chosen for the PPTA are marked by a
  star. }\label{fg:ppta_psrs}
\end{figure}

Systematic observations of the 20 MSPs chosen for the PPTA commenced
about one year ago. Fig.~\ref{fg:ppta_psrs} shows the sky distribution
of these and other MSPs with periods less than 20 ms (excuding those
in globular clusters). PPTA pulsars are generally those with
relatively narrow pulses and high flux densities as these give the
most precise TOAs, but in some cases pulsars have been included to
improve the sky distribution. This figure highlights the fact that
there are few suitable MSPs known north of the Parkes northern
declination limit ($+24\degr$). Further searches of the northern sky
for MSPs are clearly warranted. Additional searches of the southern
sky at high Galactic latitudes would also be useful.

\begin{figure}
\centerline{\includegraphics[width=70mm,angle=270]{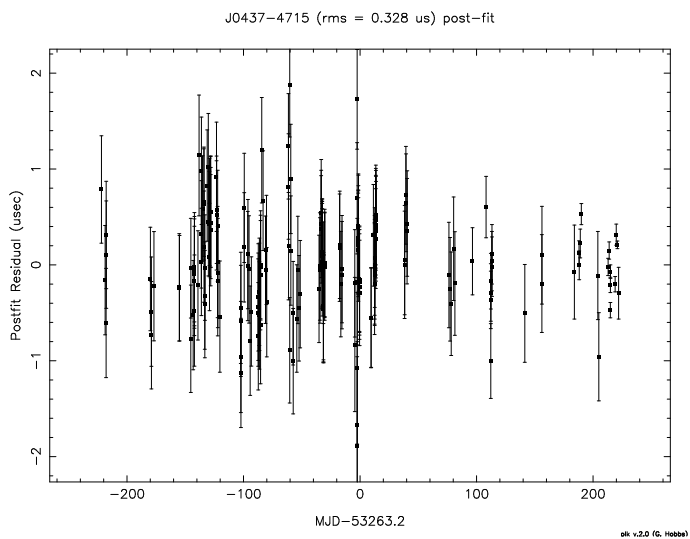}}
\caption{Timing residuals from observations of PSR J0437$-$4715 at
  3100 MHz using the wideband correlator.}\label{fg:0437res}
\end{figure}

Regular observations of the PPTA sample commenced in mid-2004, although
as discussed in \S\ref{s:goals}, development of the observing systems
has continued since then. Typically we have a 2 -- 3 day observing
session every 2 -- 3 weeks where we observe the 20 pulsars in the
sample at 700, 1400 and 3100 MHz.  CPSR2 observations at 1400 MHz are
currently giving the best results, especially when combined with
previous Swinburne timing observations. Currently we have
sub-microsecond timing residuals for seven pulsars and residuals of a
few microseconds for an additonal seven. The best timing is for PSR
J1909-3744 for which daily observations give an rms residual of 74 ns
(\cite{jhb+05}). Despite the limited time and frequency resolution of
the wideband correlator, sub-microsecond residuals have been obtained
for four pulsars at 1400 or 3100 MHz. Fig.~\ref{fg:0437res} shows
post-fit residuals to observations of PSR J0437$-$4715 at 3100 MHz
using the wideband correlator which have an rms value of 0.32 $\mu$s.
The prototype PDFB is producing high quality data and has convincingly
demonstrated the viability of the technique. However, it still has
some limitations in maximum bandwidth and minimum sampling time which
will be overcome in the final PDFB system. We anticipate that this
will be commissioned in early 2006.

\begin{figure}
\centerline{\includegraphics[width=80mm,angle=270]{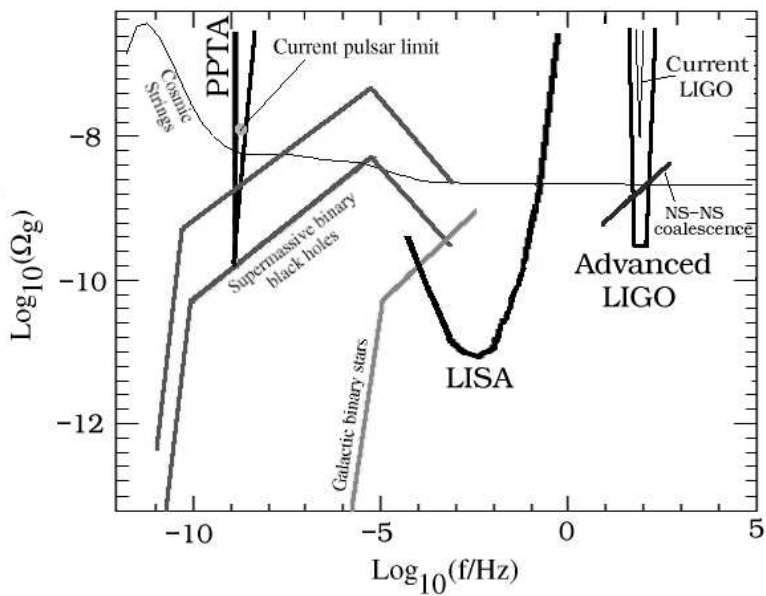}}
\caption{Energy density of gravitational waves relative to the
  critical mass density of the Universe as a function of
  frequency. Sensitivity limits of the LIGO laser interferometer
  system, both current and projected, the proposed LISA space
  interferometer system and the projected sensitivity of the PPTA
  after 5 years of observation are shown. The current limit on from
  pulsar timing and the expected spectrum from several classes of
  astronomical sources are also given. The two lines for supermassive
  binary black holes show the upper and lower bounds of the
  predictions. (After Maggiore (2000), with help from S. Phinney,
  F. Jenet and G. Hobbs)}
\label{fg:GW_spec}
\end{figure}

Simulations of the PPTA (\cite{jhlm05}) show that, with timing residuals
of 100 ns or better for at least 10 of the MSPs and 500 ns or better
for the remainder with 250 observations over 5 years, the predicted
levels of the stochastic GW background from binary black holes in
galaxies (\cite{jb03,wl03a}) should be detectable. Since the expected
timing residuals from GW have a very ``red'' spectrum, i.e., are
dominated by slow fluctuations, filtering of the data is necessary to
obtain the highest sensitivity. Optimal techniques for doing this
``pre-whitening'' are being investigated. Fig.~\ref{fg:GW_spec} shows
the sensitivity curves for LIGO (both current and Advanced LIGO,
the latter expected to be completed about 2010), LISA (proposed launch in 2012)
and the PPTA after five years, as well as the expected level of the
stochastic GW background at the Earth for a range of astronomical
sources. The current limit on the energy density from pulsar timing
(\cite{lbsn03}) is also shown. This figure highlights the
complementarity of pulsar timing to the other GW detection efforts and
shows that the PPTA should be able to detect GW from supermassive
black holes in galaxies if the technological goals are achieved. 

\nocite{mag00}

\section{Conclusion}
The Parkes Pulsar Timing Array (PPTA) project aims to obtain
high-precision timing data on a sample of 20 of the brightest
short-period MSPs accessible to the Parkes 64-m telescope. The
principal goal of the project is the direct detection of gravitational
waves with frequencies in the nanoHertz region. Simulations show that,
with a relatively modest improvement in current timing precision, this
should be possible with observations over a 5-year data span. In any
case, the data will permit a range of interesting investigations
including establishment of a pulsar-based timescale, studies of the
interstellar medium and studies of the pulsars themselves. New receiver and
software systems, including a versatile digital filterbank, an RFI
mitigation system and a timing analysis program are being developed to
help us achieve these goals.

\begin{acknowledgements}
I thank my colleagues at the ATNF and the collaborating institutions,
especially George Hobbs, Russell Edwards, Rick Jenet and Matthew
Bailes, for their major contributions to this project. The PPTA
project is supported by the Australian Research Council and the CSIRO
under the Federation Fellowship program. The Parkes telescope is part
of the Australia Telescope which is funded by the Commonwealth
Government for operation as a National Facility managed by CSIRO.

\end{acknowledgements}


\end{document}